# Application of the iterative approach to modal methods for the solution of Maxwell's equations

Igor Semenikhin[a,b,*] and Mauro Zanuccoli[c]

[a]*Institute of Physics and Technology RAS, 117218 Moscow, Russia*
[b]*Research Institute for Electrical Communication, Tohoku University, Sendai 980-8577, Japan*
[c]*ARCES-DEIS University of Bologna and IUNET, 47521 Cesena (FC), Italy*

*Abstract.* In this work we discuss the possibility to reduce the computational complexity of modal methods, i.e. methods based on eigenmodes expansion, from the third power to the second power of the number of eigenmodes. The proposed approach is based on the calculation of the eigenmodes part by part by using shift-and-invert iterative technique and by applying the iterative approach to solve linear equations to compute eigenmodes expansion coefficients. As practical implementation, the iterative modal methods based on polynomials and trigonometric functions as well as on finite-difference scheme are developed. Alternatives to the scattering matrix (S-matrix) technique which are based on pure iterative or mixed direct–iterative approaches allowing to markedly reduce the number of required numerical operations are discussed. Additionally, the possibility of diminishing the memory demand of the whole algorithm from second to first power of the number of modes by implementing the iterative approach is demonstrated. This allows to carry out calculations up to hundreds of thousands eigenmodes without using a supercomputer.

## 1. Introduction

Over the past decades an increasing interest has been devoted to the rigorous solution of Maxwell's equations in order to study the optical properties of optoelectronic devices including image sensors, nanostructured solar cells, photonic crystals and diffraction gratings [1,2]. Due to a rapid increase in complexity of the simulated devices and the continuously rising demand of accuracy that leads to the need of solving very large systems of linear equations, the application of iterative techniques [3,4] to the solution of Maxwell's equations in frequency domain may be very promising.

In recent years a number of algorithms to solve Maxwell's equations within iterative approach have been proposed [5-10]. Its implementation can significantly reduce the computational time. For example, iterative implementation of Generalized Source method, differential and Green's function approaches can reduce the number of arithmetic operations from cubic to linear dependence on the number of Fourier modes [5-7].

In case of modal methods iterative approaches are also exploited to speed up calculations, for instance to solve eigenvalue problems using Arnoldi method [8-10]. However, the computational complexity of these approaches is of order of the third power of the number of used eigenmodes, which is the same of the direct approach. The additional difficulty is due to the fact that modal methods include two significantly different stages: solution of the eigenvalue problem and calculation of the expansion coefficients with which eigenmodes represent the electromagnetic field. To decrease the order of computational complexity of the entire algorithm, both of these stages should be appropriately reformulated.

As it is well known according to modal methods, i.e. the algorithms based on eigenmode expansion, the simulation domain is divided into layers featuring parallel interfaces. The permittivity varies only in the plane of the layer and is assumed to be constant along the perpendicular direction allowing the separation of spatial variables (Fig. 1). Within each layer the eigenmodes of the electromagnetic field are calculated and the general solution is then expressed by means of an eigenmode expansion. The expansion coefficients can be found by applying proper boundary conditions. Due to their relative simplicity, two-dimensional (2-D) structures are often considered in order to discuss potentialities, drawbacks and features of the proposed algorithms. Indeed, in the case of 2-D structures, for which the dielectric constant within each parallel layer depends only on one spatial coordinate, in order to obtain the eigenmodes the one-dimensional Helmholtz equation has to be solved.

Most proposed implementations of the modal method use a set of predetermined basis functions to express the eigenmodes such as the complex Fourier series used in the Fourier modal method (FMM) known also as Rigorous Coupled-Wave Analysis method (RCWA) [11], orthogonal polynomials adopted in polynomial expansion modal methods [12-14] or B – splines [15,16]. The eigenmodes may be computed also by finite-difference approach [17] or by pseudospectral methods [18,19]. It is worth noring that all these methods are developed for periodic structures, however, using perfectly matched layers (PML) [20] they can be reformulated for implementing a general aperiodic case. From this point of view the FMM (RCWA) is especially thoroughly researched [21,22]. The direct solution of the eigenvalue problem in any of these methods demands an order of $M^3$ operations, where $M$ is the number of eigenmodes included to computation. The only exception is the analytic or true modal method (AMM) [23,24] in which eigenmodes are expressed in an exact analytical form requiring order of operations equal to $M$ to compute the eigenvalues by finding the roots of the transcendental equation. This represents a complicated task especially in case of complex permittivity and often one of the above-listed methods are used to find the approximation to the roots with subsequent purification by Newton's method [25]. In case of transverse electric polarization, a simple algorithm based on perturbation theory can be implemented [25] to avoid the solution of the full eigenvalue problem. However, for transverse magnetic polarization such simple technique is absent. Another drawback of AMM is that it can be rigorously implemented only in the 2-D case.

The computation of the expansion coefficients by S–matrix or similar technique also requires about $M^3$ operations [12,26-28], consuming a noticeable time in overall calculation.

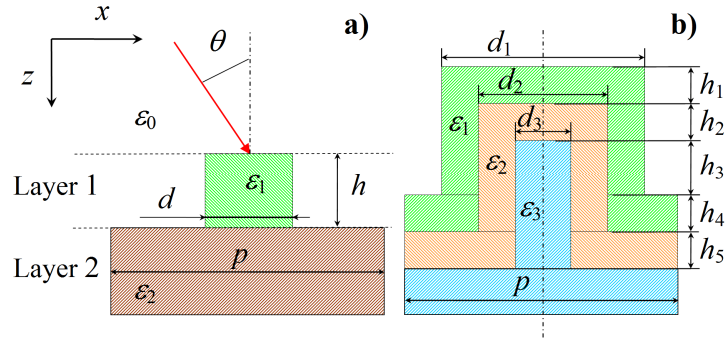

Fig. 1. Sketch of the gratings configuration. One period is depicted.

In this work we have proposed the application of the iterative approach to modal methods based on two relevant enhancements that can significantly increase the efficiency of the algorithms:
 1) the $M$ eigenmodes are found *by parts* by using Arnoldi method and shift-and-invert technique;
 2) in order to solve the system of linear equations an *iterative* procedure is used instead of a S or a R-matrix formalism.
If the possibility to restrict the number of operations in one shift-and-invert iteration to a value proportional to the first power of the number of used basis functions (or mesh points in pseudospectral and finite-difference methods) exists, then in combination to point 1) this ensure a reduction in a number of arithmetic operations to the order of $M^2$ to solve the eigenvalue problem. In some cases it can be obtained by using Fast Fourier transform (FFT), in others by using some features of considered matrix such as sparseness or band structure. The point 2) allows to minimize the computational time of determining the expansion coefficients. Additionally, the memory requirements are relaxed from $M^2$ to the order of $M$ in some formulations of the algorithm.

The iterative modal methods formulated within proposed approach may have its already known direct counterpart or have not. In the next section we present two examples of application of this approach, one of which use Chebyshev polynomials as basis function and has its direct counterpart [18], the over based on trigonometric functions has not.

The remainder of the paper is organized as follows: in Section 2 the theoretical part concerning the 2-D implementation case is outlined. In particular in section "A" we describe the iterative solution of eigenvalue problem and in section "B" the calculation of the expansion coefficients by iterative methods. In addition, the implementation of introduced techniques to three-dimensional (3-D) case is outlined. In the third part of the paper we discuss the accuracy and the computational efficiency of the proposed approach. A comparison with the classical FMM (RCWA) and FMM with adaptive special resolution or parametric FMM (PFMM) [29], which are very popular and widely used methods, is provided.

## 2. Implementation

The general solution of Maxwell's equations for nonconical diffraction in the 2-D case can be represented as a linear combination of two fundamental polarizations: the transverse electric polarization (TE) and the transverse magnetic one (TM). In the case of TE polarization the electric field within each $l$-th layer ($l=0..L$, where $l=0$ is superstrate) satisfies the following equation [1]:

$$\frac{\partial^2 E_y}{\partial z^2} + \frac{\partial^2 E_y}{\partial x^2} = -k_0^2 \varepsilon_l(x) E_y, \quad E_x = E_z = 0, \quad \text{(TE)} \tag{1}$$

where $\varepsilon_l(x)$ is the relative dielectric constant of the $l$-th layer, $k_0 = 2\pi/\lambda$ and $\lambda$ denotes the wavelength of the incident radiation. The relative permeability is set equal to unity. The incident wave is described by $E_y^{inc} = \exp[ik_0 n_0(x\sin\theta + z\cos\theta)]$, where $\theta$ is the angle of incidence of the radiation with respect to the normal to the device plane (Fig. 1) and $n_0$ is the refractive index of the superstrate. Along $x$ axis the quasiperiodic Floquet boundary conditions are applied:

$$E_y(x+p,z) = \exp(ik_0 n_0 p \sin\theta) E_y(x,z), \tag{2}$$

where $p$ is the periodicity of grating. After separating the variables in Eq. (1), the solution of such equation within each $l$-th layer can be written as:

$$E_y(x,z) = \sum_n \left[ c_{l,n}^+ \exp(ik_0\sqrt{\kappa_{l,n}}z) + c_{l,n}^- \exp(-ik_0\sqrt{\kappa_{l,n}}z) \right] \varphi_{l,n}(x), \tag{3}$$

where $c_{l,n}^+$ and $c_{l,n}^-$ are the amplitudes of forward and backward propagating modes, $\kappa_{l,n}$ and $\varphi_{l,n}(x)$ are the eigenvalues and the eigenfunctions of the one-dimensional Helmholtz equation:

$$\frac{d^2\varphi(x)}{k_0^2 dx^2} + \varepsilon_l(x)\varphi(x) = \kappa\varphi(x) \tag{4}$$

with boundary conditions:

$$\varphi(x+p) = \exp(ik_0 n_0 p \sin\theta)\varphi(x). \tag{5}$$

In the case of TM polarization the equation for the magnetic field at each $l$-th layer is [1]:

$$\frac{\partial^2 H_y}{\partial z^2} = -\varepsilon_l(x)\frac{\partial}{\partial x}\left(\frac{1}{\varepsilon_l(x)}\frac{\partial H_y}{\partial x}\right) - k_0^2 \varepsilon_l(x) H_y, \quad H_x = H_z = 0, \quad \text{(TM)} \tag{6}$$

with the quasiperiodic Floquet boundary conditions:

$$H_y(x+p,z) = \exp(ik_0 n_0 p \sin\theta) H_y(x,z), \tag{7}$$

and the incident wave: $H_y^{inc} = \exp[ik_0 n_0(x\sin\theta + z\cos\theta)]$. After separating the variables in Eq.(6), the solution of such equation within each $l$-th layer can be written as:

$$H_y(x,z) = \sum_n \left[ c_{l,n}^+ \exp(ik_0\sqrt{\kappa_{l,n}}z) + c_{l,n}^- \exp(-ik_0\sqrt{\kappa_{l,n}}z) \right] \psi_{l,n}(x), \tag{8}$$

where $\kappa_{l,n}$ and $\psi_{l,n}(x)$ are the eigenvalues and the eigenfunctions of the one-dimensional Helmholtz equation:

$$\varepsilon_l(x)\frac{d}{k_0 dx}\left(\frac{1}{\varepsilon_l(x)}\frac{d\psi(x)}{k_0 dx}\right) + \varepsilon_l(x)\psi(x) = \kappa\psi(x), \tag{9}$$

$$\psi(x+p) = \exp(ik_0 n_0 p \sin\theta)\psi(x). \qquad (10)$$

Eq. (4) and (9) can be written in general form as $\Lambda\psi = \kappa\psi$ where $\Lambda$ is an appropriate differential operator and $\kappa$ is corresponding eigenvalue. The sign of square root in (3) and (8) is chosen so that its imaginary part would be positive, whereas for real positive $\kappa_{l,n}$ the positive square root should be used. Thus, the directions of propagation and decaying of modes are the same.

In this paper, we mainly devote attention to TM-case due to its higher complexity. However, results for TE case are very similar and we will briefly discuss the differences in formulae.

### A. Iterative solution of eigenvalue problem

Iterative methods exploiting shift-and-invert technique are based on creation of a sequence of the type: $\psi_{i+1} = (\Lambda-\kappa)^{-1}\psi_i/\alpha$ where $\alpha$ is the value of function $\psi_i$ at the point where $|\psi_i|$ exhibits its maximum and $\kappa$ is the chosen eigenvalue shift. This sequence converges to the eigenfunction of the operator $\Lambda$ associated to the eigenvalue equal to $\kappa + 1/\alpha$ which is closest to value of shift $\kappa$ [3,30]. If $\kappa$ is changed at each step by the estimated eigenvalue, then the algorithm can converge very rapidly. A number of eigenfunctions corresponding to eigenvalues which are closest to eigenvalue shift $\kappa$ can be found simultaneously if the Arnoldi iteration algorithm [31] with operator $(\Lambda-\kappa)^{-1}$ is implemented [3,30]. By changing consecutively the value of shift $\kappa$, the entire set of *M* required eigenmodes of operator $\Lambda$ can be calculated part by part.

Two different approaches within modal method are possible. Following the first one, the electromagnetic field is represented by a complete set of eigenfunctions of matrix that approximates operator $\Lambda$ in some basis. This approach is used for example in Fourier modal method, finite-difference modal method [17], PEMM [12], MMGE [13,14], PSMM [19], and in most methods mentioned above. Due to the finite size of the matrix, usually only small part of its eigenfunctions approximates true eigenfunctions of operator $\Lambda$ with good accuracy. The second approach consists in retaining only such accurate eigenfunctions neglecting the others. Examples of these methods are those described in [8,9], IFDMM [10], and PSMM [18]. The accuracy of these methods tends to that of AMM and is limited by accuracy of AMM. From this point of view they can be classified as AMM-like methods. At first we consider this second approach in terms of methods based on spectral elements: iterative polynomial expansion modal method (IPEMM) and iterative trigonometric expansion modal method (ITEMM).

To implement the iterative approach in the solution of the eigenvalue problem (4)-(5) and (9)-(10) we need to solve equations of the form:

$$(\Lambda - \kappa)\psi(x) = f(x), \qquad (11)$$

with boundary condition (5) and (10), respectively, for any given function *f* and value of $\kappa$. With this regard the *l*-th layer is divided into *J* elements along *x*-axis, in each of which the permittivity is spatially uniform, and for each *j*-th element (*j*=1..*J*) the mesh with nodes $x_i^{l,j}$, $i = 0..N^{l,j}$ is used. The total number of nodes is $N^l = \sum_{j=1,J} N^{l,j}$. The adopted mesh is uniform for ITEMM and nonuniform for IPEMM. In the following of this paragraph we will omit the index of layer *l* for conciseness.

In the spectral element method the solution of Eq. (11) within every *j*-th element can be conveniently written in the form [32]:

$$\psi_j(x) = \psi_j^0(x) + b_j^L \psi_j^L(x) + b_j^R \psi_j^R(x), \qquad (12)$$

where $b_j^L$, $b_j^R$ are unknown coefficients and $\psi_j^0(x)$ is the solution of Eq.(11) with zero boundary conditions, $\psi_j^L(x)$ and $\psi_j^R(x)$ are solutions of the homogeneous equation corresponding to (11) with conditions: $\psi_j^L(x_0^j) = 1$, $\psi_j^R(x_0^j) = 0$ and $\psi_j^L(x_{N^j}^j) = 0$, $\psi_j^R(x_{N^j}^j) = 1$ at left and right boundaries of the element respectively. It is worth noting that if we substitute the exact eigenvalue and the corresponding eigenfunction to (11), the function *f(x)* becomes identically zero, therefore $\psi_j^0(x)$ in the expansion of the exact eigenfunction is also identically zero, as a solution of the homogeneous linear differential equation with zero boundary conditions.

Coefficients $b_j^L$, $b_j^R$ can be found by applying proper matching conditions at the boundaries of elements. From the condition describing the continuity $\psi(x)$ at the boundaries of elements and from the quasiperiodic conditions (5) for TE and (10) for TM cases:

$$b_j^R = b_{j+1}^L, \ (j = 1, J-1), \ b_J^R = \exp(i k_0 n_0 p \sin \theta) b_1^L. \tag{13}$$

And from continuity of $\varepsilon^{-1} d\psi/dx$ for TM-case we obtain:

$$\varepsilon_j^{-1} \frac{d}{dx}\left(\psi_j^0 + b_j^L \psi_j^L + b_j^R \psi_j^R\right)\Big|_{x=x_N^j} = \varepsilon_{j+1}^{-1} \frac{d}{dx}\left(\psi_{j+1}^0 + b_{j+1}^L \psi_{j+1}^L + b_{j+1}^R \psi_{j+1}^R\right)\Big|_{x=x_0^{j+1}}, (j = 1, J-1)$$

$$\varepsilon_J^{-1} \frac{d}{dx}\left(\psi_J^0 + b_J^L \psi_J^L + b_J^R \psi_J^R\right)\Big|_{x=x_N^J} = \exp(i k_0 n_0 p \sin\theta) \varepsilon_1^{-1} \frac{d}{dx}\left(\psi_1^0 + b_1^L \psi_1^L + b_1^R \psi_1^R\right)\Big|_{x=x_0^1} \tag{14}$$

In case of TE polarization the continuity condition of $d\varphi/dx$ must be satisfied and there are no permittivities in Eq. (14). After calculating of the derivatives $d\psi_j/dx$ and by substituting (13) to (14) we obtain the system of linear equations with cyclic-tridiagonal matrix which can be solved in order of $J$ arithmetic operations.

The functions $\psi_j^L, \psi_j^R$ can be expressed analytically. The general solution of the homogeneous equation is:

$$\psi_j^{L(R)} = \alpha_j^{L(R)} \exp(\gamma_j x) + \beta_j^{L(R)} \exp(-\gamma_j x), \tag{15}$$

where $\gamma_j = ik_0(\varepsilon_j - \kappa)^{1/2}$ and coefficients $\alpha_j^L$, $\beta_j^L$ are found from conditions on boundaries of element:

$$\alpha_j^L = \exp(\gamma_j x_0^j) / \left[\exp(2\gamma_j x_0^j) - \exp(2\gamma_j x_N^j)\right]$$
$$\beta_j^L = \exp(-\gamma_j x_0^j) / \left[\exp(-2\gamma_j x_0^j) - \exp(-2\gamma_j x_N^j)\right]. \tag{16}$$

The expressions for coefficients $\alpha_j^R$, $\beta_j^R$ differ from (16) only by a change in coordinate $x_0^j$ by $x_N^j$ and vice versa.

Functions $\psi_j^0$ can be calculated either by polynomial (Appendix A) or by trigonometric expansion method (Appendix B). Alternatively, functions $\psi_j^0$ can be calculated by any other method that requires an order of $N^{l,j}$ operations to solve Eq. (11) with zero boundary conditions inside the *j*-th element. Throughout Arnoldi iterations the vectors of functions values (12) at mesh points are used. In order to convert $\psi_j^0(x)$ and *f*(x) from values at mesh points to Chebyshev polynomials (or trigonometric functions) expansion and vice-versa, FFT is implemented (see Appendixes A and B). Thus, the total number of arithmetic operations required to find solutions of (11) in layer *l* is of order of *N*. Consequently, finding simultaneously *m* eigenvectors by Arnoldi iterations [3] takes an order of $m^2 N$ operations, where *m* is small enough ($m \ll N$), usually about several dozens. The set of *M* eigenmodes (*M≤N*) is found by order of $MNm \sim M^2$ operations.

The whole algorithm of iterative solution of the eigenvalue problem consists of repeating two stages:
1) choosing the eigenvalue shift $\kappa$ and finding *m* eigenfunctions corresponding to those closest to $\kappa$ eigenvalues by Arnoldi iterations;
2) purifying each eigenfunction and eigenvalue by means of shifted inverse iterations.

The shift $\kappa$ should be chosen in order to produce an overlap between the already calculated and the new part of eigenvalues to prevent eigenvalue missing. The value of $\kappa$ can be estimated by taking into account intervals between the latest already found eigenvalues. Better accuracy is achieved by comparing these eigenvalues with these of an uniform layer that can be found analytically. The purification of eigenmodes previously found by Arnoldi method allows to attain the relative accuracy of eigenfunctions and eigenvalues close to machine precision when analytic expressions (15) are used. The reason of this excellent accuracy is that the norm of function $\psi_j^0(x)$ goes to zero for exact eigenvalue, as observed above. Therefore, relative inaccuracies in function $\psi_j^0(x)$ do not affect significantly the final result. Furthermore, the eigenvalues can be additionally purified by Newton's method as discussed in [25].

The eigenfunction corresponding to the eigenvalue $\kappa$ in each element $j$ can be expressed in the following exact analytic form:

$$\psi_j = \alpha_j \exp(\gamma_j x) + \beta_j \exp(-\gamma_j x), \tag{17}$$

where $\gamma_j = ik_0(\varepsilon_j - \kappa)^{1/2}$ [23,24]. Therefore, instead of storing in the computer memory the eigenfunctions values in mesh points, it is possible to conveniently store only coefficients $\alpha_j$, $\beta_j$ and $\gamma_j$. This significantly reduces memory requirements and allows to calculate matrix elements that are needed subsequently by analytical integration with order of $M^2$ operations. The coefficients $\alpha_j$ and $\beta_j$ can be calculated, for example, from values of eigenfunction at boundaries of $j$-th element. Similarly, the eigenfunctions for known eigenvalues can be found in the form of (17) by order of $M$ operations by posing proper matching conditions at the boundaries of elements as discussed in [25] for AMM or from Eq. (13)-(14) in addition to condition $\psi_j^0 \equiv 0$.

The eigenfunctions of Eq. (4) and (9) can also be found without representing them in the form of Eq. (12). In case of IPEMM the solution of (11) is written as $\psi_j(\xi) = \sum_{k=0}^{N_j} d_k T_k(\xi)$ instead of sum (15) and (32). The coefficients $d_k$ are found in a way similar to that described in appendix A, whereas values of eigenfunction at boundaries can be calculated for instance by generalized minimal residual method (GMRES) [4] from matching conditions. Similarly for ITEMM, with the only difference that, in this case, the auxiliary function is found in the form $\vartheta(\xi) = \sum_{n=0}^{K} a_n \cos n\xi$ instead of (47). In case of IPEMM such iteration approach is entirely equivalent to its direct counterpart described in [18] and in theory can be used to find complete set of eigenfunction to apply the technique discussed in [19]. However, up to now we have not achieved sufficient accuracy of the eigenvalues with largest magnitudes by means of the iterative approach and in the following we limit the discussion only to AMM-like methods.

### B. Calculation of the expansion coefficients by iterative method

The firsts $M$ eigenstates are calculated within each of the $l$-th layer of $L$, then the expansion coefficients $c_{l,n}^+$, $c_{l,n}^-$ are determined by imposing the continuity conditions for $H_y(x,z)$ and $\varepsilon^{-1} \partial H_y(x,z)/\partial z$ (TM-case) at the boundary $z_l$ between $(l-1)$-th and $l$-th layers:

$$H_y\big|_{z=z_l-0} = H_y\big|_{z=z_l+0}, \quad \frac{1}{\varepsilon}\frac{\partial H_y}{\partial z}\bigg|_{z=z_l-0} = \frac{1}{\varepsilon}\frac{\partial H_y}{\partial z}\bigg|_{z=z_l+0}. \tag{18}$$

For the TE-case the continuity conditions for $E_y(x,z)$ and $\partial E_y(x,z)/\partial z$ are used instead. By substituting $H_y(x,z)$ with its eigenmodes expansion (8), by multiplying by test functions $\{\phi_m(x)\}_{m=1..M}$ and lastly by integrating over the period $p$, we obtain a system of $L \cdot 2M$ linear equations in order to determine the expansion coefficients $c_{l,m}^+$, $c_{l,m}^-$:

$$\begin{aligned}
\mathbf{A}_{l-1}\left[\exp(\mathbf{\eta}_{l-1})\mathbf{c}_{l-1}^+ + \exp(-\mathbf{\eta}_{l-1})\mathbf{c}_{l-1}^-\right] &= \mathbf{A}_l\left[\mathbf{c}_l^+ + \mathbf{c}_l^-\right] \\
\mathbf{B}_{l-1}\left[\exp(\mathbf{\eta}_{l-1})\mathbf{c}_{l-1}^+ - \exp(-\mathbf{\eta}_{l-1})\mathbf{c}_{l-1}^-\right] &= \mathbf{B}_l\left[\mathbf{c}_l^+ - \mathbf{c}_l^-\right]
\end{aligned}, \tag{19}$$

where $\mathbf{\eta}_l = ik_0 \mathbf{\kappa}_l^{1/2} h_l$, $h_l = z_l - z_{l-1}$ is a diagonal matrix and elements of matrix $A_l$, $B_l$ are:

$$\left(\mathbf{A}_l\right)_{n,m} = \int_0^p \phi_n^*(x)\psi_{l,m}(x)dx, \quad \left(\mathbf{B}_l\right)_{n,m} = \kappa_{l,m}^{1/2}\int_0^p \phi_n^*(x)\varepsilon_l^{-1}(x)\psi_{l,m}(x)dx. \tag{20}$$

If $\delta$ denotes the Kronecker delta, the coefficients $c_{0,m}^+ = \delta(0,m)$, and $\mathbf{c}_L^- \equiv 0$ are known and specify the zero order incident wave in superstrate and the upward going wave in substrate, respectively. As test functions the $M$ firsts eigenfunctions of the superstrate layer can be implemented. This is equivalent to Fourier matching scheme.

Usually, the system (19) is solved by means of S-matrix approach [28]. However, such system can be solved by using any direct band matrix solver by implementing only small changes within the system. It is worth noting that rows in system (19) include exponentially large and exponentially small values simultaneously. This leads to numerical instability when $h_l / p$ is large or many eigenmodes are included in the calculation. This instability can be avoided simply by defining new variables:

$$c_{l,n}^- = \tilde{c}_{l,n}^- \exp(\eta_{l,n}), \tag{21}$$

where the multiplier $\exp(\eta_{l,n})$ is relatively small due to particular choice of the sign of the imaginary part of $\kappa_l^{1/2}$ (Im($\kappa_l^{1/2}$)>0). Physically this means that the expansion coefficients of upward and downward waves are defined at opposite sides of the layer. In addition, the determination of $\mathbf{c}^+$ and of $\tilde{\mathbf{c}}^-$ opens the way to a straightforward and stable calculation of the field within each layer avoiding computationally-intensive operations with partial S-matrices such as those proposed in [26]. By using the variables (21), exponentially large numbers are avoided in the system:

$$\begin{aligned}\mathbf{A}_{l-1}\left[\exp(\mathbf{\eta}_{l-1})\mathbf{c}_{l-1}^+ + \tilde{\mathbf{c}}_{l-1}^-\right] &= \mathbf{A}_l\left[\mathbf{c}_l^+ + \exp(\mathbf{\eta}_l)\tilde{\mathbf{c}}_l^-\right] \\ \mathbf{B}_{l-1}\left[\exp(\mathbf{\eta}_{l-1})\mathbf{c}_{l-1}^+ - \tilde{\mathbf{c}}_{l-1}^-\right] &= \mathbf{B}_l\left[\mathbf{c}_l^+ - \exp(\mathbf{\eta}_l)\tilde{\mathbf{c}}_l^-\right]\end{aligned}, \tag{22}$$

and it can be solved for instance by means of block Gauss elimination requiring 3 matrix-matrix multiplications and 2 solutions of linear equations systems with $M$ right-hand sides per each internal layer. For the two outer layers the equations are simpler since the coefficients of ongoing waves are known. For instance the structure of the matrix of system (22) for TM-case for three layers is:

$$\begin{pmatrix} * & * & e^- & & & & \\ * & * & e^- & & & & \\ e^- & * & * & e^- & & & \\ e^- & * & * & e^- & & & \\ & & & & e^- & * & * \\ & & & & e^- & * & * \end{pmatrix}\begin{pmatrix} \tilde{\mathbf{c}}_0^- \\ \mathbf{c}_1^+ \\ \tilde{\mathbf{c}}_1^- \\ \mathbf{c}_2^+ \\ \tilde{\mathbf{c}}_2^- \\ \mathbf{c}_3^+ \end{pmatrix} = \begin{pmatrix} \mathbf{c}_0^+ \\ -\kappa_0^{1/2}\varepsilon_0^{-1}\mathbf{c}_0^+ \\ 0 \\ 0 \\ 0 \\ 0 \end{pmatrix}, \tag{23}$$

where $e^-$ indicates the matrices with decreasing exponential multiplier. For TE-case the inverse permittivity $\varepsilon_0^{-1}$ in the right hand side of (23) is not present.

The operation count can be reduced if adjoint eigenfunctions $\psi_{l,n}^+$ are implemented as test functions. The functions $\psi_{l,n}^+$ satisfy differential operator (9) (or (4) for TE–case) with complex conjugated permittivity $\varepsilon_l^*(x)$ [24]. Such functions correspond to known complex conjugated eigenvalues $\kappa_{l,n}^*$, therefore they can be found by order of $M$ arithmetic operations only, as discussed above. By using $\psi_{l,n}^+$ as test functions, the linear equations (22) for inner layers $l=2..L$-1 can be rewritten as:

$$\begin{aligned}\exp(\mathbf{\eta}_{l-1})\mathbf{c}_{l-1}^+ + \tilde{\mathbf{c}}_{l-1}^- &= \tilde{\mathbf{A}}_l\left[\mathbf{c}_l^+ + \exp(\mathbf{\eta}_l)\tilde{\mathbf{c}}_l^-\right] \\ \kappa_{l-1}^{1/2}\left[\exp(\mathbf{\eta}_{l-1})\mathbf{c}_{l-1}^+ - \tilde{\mathbf{c}}_{l-1}^-\right] &= \tilde{\mathbf{B}}_l\left[\mathbf{c}_l^+ - \exp(\mathbf{\eta}_l)\tilde{\mathbf{c}}_l^-\right]\end{aligned}, \tag{24}$$

where:

$$\left(\tilde{\mathbf{A}}_l\right)_{n,m} = \int_0^p \psi_{l-1,n}^{+*}(x)\varepsilon_{l-1}^{-1}(x)\psi_{l,m}(x)dx, \quad \left(\tilde{\mathbf{B}}_l\right)_{n,m} = \kappa_{l,m}^{1/2}\int_0^p \psi_{l-1,n}^{+*}(x)\varepsilon_l^{-1}(x)\psi_{l,m}(x)dx, \tag{25}$$

and the property of orthogonality of $\psi_{l,n}^+$ and $\psi_{l,m}$ with weight $\varepsilon_l^{-1}(x)$ is used for TM-polarization (for TE case the weight function is unity [23,24]). The system (24) exhibits the same structure of (23) but with different matrix elements. To solve such system by block Gauss eliminations, again only two solutions of systems of linear equations with $M+1$ right-hand sides are required per layer.

An additional reduction in operation count is possible by implementing iteration techniques. Everywhere in this work the GMRES method [4] is implemented as iterative solver where results of matrix Eq. (23) multiplication by vectors are needed as an input. To accelerate the convergence of iterations a preconditioner is used. The preconditioner matrix should be as close as possible to the matrix of the equation to solve and at the same time it should be easily invertible. The choice of preconditioner is the most critical in iterative methods; an effective preconditioner can dramatically reduce the number of iterations. One of the simplest and most effective option in our case is to use the part of the matrix of Eq. (23) marked by asterisks, neglecting the submatrices with decreasing exponential multiplier. The blocks of such matrix features the form:

$$\begin{pmatrix} \mathbf{I} & -\tilde{\mathbf{A}}_l \\ -\boldsymbol{\kappa}_{l-1}^{1/2} & -\tilde{\mathbf{B}}_l \end{pmatrix} \quad (26)$$

and to solve the corresponding linear system the calculation of one inverse matrix $(\boldsymbol{\kappa}_{l-1}^{1/2}\tilde{\mathbf{A}}_l + \tilde{\mathbf{B}}_l)$ is enough. After performing the $LU$ factorization of such matrix, the subsequent solutions of preconditioner system are calculated by order of $M^2$ operations per iteration. The average number of iterations in our calculations with this preconditioner to achieve good accuracy is usually proportional to the number of layers multiplied by a factor ranging from 3 to 10. Such factor depends on the value of the permittivity and on the geometry of the layers which define the magnitude of rejected submatrices with decreasing exponential multiplier. The overall number of operations required to solve system (24) is about of $2/3\,M^3$ per layer if the operations which are of the order of $M^2$ are neglected. To further reduce the computational time, the $LU$ factorization for preconditioner can be carried out in single precision.

Note that if calculations of $\tilde{\mathbf{A}}_l, \tilde{\mathbf{B}}_l$ are not possible analytically by Eq. (25) as in case of direct methods such as FMM, they can be calculated numerically by $5/3M^3$ operations for each matrix by calculating the $LU$ decompositions of $\mathbf{A}_{l-1}$ and $\mathbf{B}_{l-1}$ at left side of Eq.(22).

Although the approaches described above allow to significantly decrease the number of operations, they are still of the order of $M^3$. It is possible to attain the numerical complexity of the order of $M^2$ by means of small variations of the preconditioner and by exploiting the properties of adjoint eigenfunctions. With this regard, we choose as test functions the eigenfunctions of the superstrate $\psi_{0,m}, m=1..M$ (Fourier matching scheme). We define the preconditioner matrix as the part of matrix of Eq. (23) that is marked by asterisks but without one lower submatrix. Each block of this preconditioner matrix and the corresponding part of the linear system are:

$$\begin{pmatrix} \mathbf{A}_{l-1} & -\mathbf{A}_l \\ 0 & -\mathbf{B}_l \end{pmatrix} \Leftrightarrow \begin{matrix} \mathbf{A}_{l-1}\tilde{\mathbf{c}}_{l-1}^- - \mathbf{A}_l\mathbf{c}_l^+ = \mathbf{u}_l \\ -\mathbf{B}_l\mathbf{c}_l^+ = \mathbf{v}_l \end{matrix}, \quad (27)$$

where $\mathbf{u}_l, \mathbf{v}_l$ are intermediate vectors that appear in the process of GMRES iterations with preconditioner, and submatrices $\mathbf{A}_l$, $\mathbf{B}_l$ are defined in (20). This linear system can be rearranged in the following equivalent form:

$$\begin{matrix} \boldsymbol{\psi}_{l-1}\tilde{\mathbf{c}}_{l-1}^- - \boldsymbol{\psi}_l\mathbf{c}_l^+ = \boldsymbol{\psi}_0\mathbf{u}_l + \xi_1 \\ -\varepsilon_l^{-1}\boldsymbol{\psi}_l\boldsymbol{\kappa}_l^{1/2}\mathbf{c}_l^+ = \boldsymbol{\psi}_0\mathbf{v}_l + \xi_2 \end{matrix}, \quad (28)$$

where $\boldsymbol{\psi}_l$ is the row of eigenfunctions of the form: $\{\psi_{l,1}(x),...,\psi_{l,M}(x)\}$ and $\xi_{1,2}(x)$ are some unknown functions that are orthogonal to all test functions $\psi_{0,m}, m=1..M$. The norm of $\xi_{1,2}$ in space $L^2$ with increasing $M$ should go to zero. By multiplying (28) by column of complex conjugate adjoint eigenfunctions $\psi_{l,n}^{+*}$ with weight function $\varepsilon_l^{-1}(x)$ for TM-case (or unity for TE), and by integrating over the period, the following approximation of the system (27) is obtained:

$$\begin{matrix} \tilde{\mathbf{c}}_{l-1}^- - \tilde{\mathbf{A}}_l\mathbf{c}_l^+ \approx \mathbf{A}_{l-1}^{inv}\mathbf{u}_l \\ \mathbf{c}_l^+ \approx -\mathbf{B}_l^{inv}\mathbf{v}_l \end{matrix} \quad (29)$$

where $\mathbf{A}_{l-1}^{inv}, \mathbf{B}_l^{inv}$ are approximation of inverse matrices $\mathbf{A}_{l-1}^{-1}, \mathbf{B}_l^{-1}$:

$$\left(\mathbf{A}_l^{inv}\right)_{n,m} = \int_0^p \psi_{l,n}^{+*}(x)\varepsilon_l^{-1}(x)\psi_{0,m}(x)dx, \quad \left(\mathbf{B}_l^{inv}\right)_{n,m} = \kappa_{l,n}^{-1/2}\int_0^p \psi_{l,n}^{+*}(x)\psi_{0,m}(x)dx, \quad (30)$$

and inner products $\psi_{l,n}^{+*}$ with $\xi_{1,2}$ are neglected. All inner products (30) are calculated analytically as in previous cases by order of $M^2$ operations only. Therefore, the elements of these matrices can be permanently calculated through the iteration process without storing them in memory. This allows a decrease in memory demand of the whole algorithm from second to first power of the number of modes. It is worth noting that it is not necessary to calculate the matrix $\tilde{\mathbf{A}}_l$ in Eq. (29) because after the $\mathbf{c}_l^+$ calculation, coefficients $\tilde{\mathbf{c}}_{l-1}^-$ are found from equation $\tilde{\mathbf{c}}_{l-1}^- \approx \mathbf{A}_{l-1}^{inv}(\mathbf{u}_l - \mathbf{A}_l\mathbf{c}_l^+)$. Sometimes, when layer $l$ is uniform, the value of $\kappa_{l,n}$ in (30) can be exactly zero for certain $n$. In this case another block for preconditioner matrix can be chosen as zero instead of the bottom left term in Eq. (27). Also in case of an uniform layer, a better preconditioner for two adjacent layers at once can be constructed by additionally considering few submatrices marked in Eq. (23) as $e^-$.

The average number of iterations in our calculations with preconditioner (27) to achieve good accuracy usually is about 30-50 per layer depending on permittivity and on geometry of the layers. If the calculation of $\mathbf{A}_l^{inv}$ and of $\mathbf{B}_l^{inv}$ is not possible analytically as in case of direct methods such as FMM, then the *LU* factorization of $\mathbf{A}_l, \mathbf{B}_l$ can be performed numerically by $2/3M^3$ operations for each matrix. In case of real permittivity the adjoint eigenfunctions $\psi_{l,n}^+$ are equal to $\psi_{l,n}$ [24] and the matrix $\mathbf{A}_l^{inv}$ and $\mathbf{B}_l^{inv}$ can be directly found:

$$\left(\mathbf{A}_l^{inv}\right)_{n,m} = \left(\mathbf{B}_l\right)_{m,n}^* \kappa_{l,n}^{-1/2}, \quad \left(\mathbf{B}_l^{inv}\right)_{n,m} = \left(\mathbf{A}_l\right)_{m,n}^* \kappa_{l,n}^{-1/2}. \quad (31)$$

Therefore, by using this approach in case of dielectric gratings direct methods such as FMM consume only of an order of $M^2$ numerical operations to find the expansion coefficients [33].

**Remarks concerning implementation of proposed approach in three dimensional (3D) case: an outline.**

In this paper we mainly discuss the implementations of iterative approach to modal methods in 2D case. Nevertheless the same techniques can be applied also to 3D spatial domains. Similarly to the 2D case, the 3D simulation domain is divided into 2D layers featuring parallel interfaces where the permittivity varies only in the plane of the layer and is assumed to be constant along the perpendicular direction. The eigenmodes within each layer can be calculated by parts by solving a 2D Helmholtz-like equation, and the system of linear equations (22) can be solved by iterative methods to find expansion coefficients. The main difference is that there are no direct economizing schemes for solving 2D Eq. (11) by order of $M$ operations in general case. In order to preserve the second power complexity of the number of modes, Eq. (11) must be solved by an appropriate iterative technique. Also in case of 2D Helmholtz-like equations, the implementation of spectral elements method is much more complicated and analytical representation of eigenmodes is not possible in the general case. The implementation of finite-difference modal method in this case is more convenient and simple since the difference between solving 1D and 2D Helmholtz-like equations within this approach is minimal in comparison with other methods. Also the calculations of all matrices within IFDMM in Eq. (22) and Eq. (30) are possible by order of $N^2 \log N$ operations by means of Fast Fourier Transform. Therefore, realizations of this method in 3D can maintain complexity about of order of $M^2$. In this paper we include for comparisons the IFDMM method [10] based on iterative solution of Eq. (11) by the GMRES solver. In this way, the computation efficiency of the corresponding 3D implementation can be roughly estimated.

**3. Numerical Results**

In order to calculate eigenmodes in the proposed numerical implementation of the algorithms, we utilize both the ARPACK library [34] and an in-house subprogram (also based on the Arnoldi method) where we reuse part of already found eigenmodes to construct the orthonormal basis of the Krylov subspace for calculating the next portion of the eigenvectors. This approach allows increasing performance by half. In all subsequent tests an appropriate subprogram finds eigenmodes part by part calculating about 20 eigenmodes at once in IPEMM and ITEMM and about 30 eigenmodes in IFDMM methods. The number of mesh points $N$ is twice larger than the number of modes $M$ for spectral element based methods IPEMM and ITEMM. Whereas, in the case of IFDMM method, in which the division by elements is not implemented, the mesh points are chosen as discussed in Ref. [10]. The average number of iterations per eigenmode calculated by Arnoldi method followed by purification by shift-and-invert power technique to achieve relative precision better than $10^{-13}$ is about 3-5 for IPEMM and 4-6 for ITEMM methods. In the case of IFDMM with iterative solution of Eq. (11) by GMRES algorithm the purification is not used. For all examples, in IFDMM we choose the number of mesh nodes $N$ to place mesh points at the permittivity discontinuities where the value of one of nearby continuous part without the averaging is assigned. As FMM (RCWA) and PFMM we implement the algorithm proposed in [35] and [29] respectively. The value of the adaptive special resolution parameter $\eta$ in PFMM is chosen to reach best convergence.

The first example is the calculation of the absorption of the metallic grating Fig. 1a) with parameters: $p=1\mu m$, $h=p$, $d=p/2$, $\varepsilon_0=1$, $\varepsilon_1=\varepsilon_2= (0.22+6.71i)^2$. The wavelength $\lambda$ of the TM polarized incident field forming an angle of incidence $\theta=30^0$ is equal to 1 $\mu m$. This grating has been previously examined in [10,13,15-19].

It is interesting to note that in spectral element based methods IPEMM and ITEMM when analytical expressions (15) are used, the convergence rate of first eigenvalues that usually are discussed in papers actually cannot be demonstrated, as distinguishable from FMM or IFDMM [10]. Only few points in each $j$-th spectral element are enough to achieve accuracy of the first eigenvalue with maximum real part close to machine precision, i.e. smaller then $10^{-15}$. The reason of such behavior is that the norm of the function $\psi_j^0(x)$ which is calculated by spectral method goes to zero for exact eigenvalue and does not contribute appreciably to the result, as pointed out above.

In Fig. 2 the computational time of the eigenmodes calculation versus number of modes for Layer 1 of the discussed metallic grating is illustrated. Since the method of calculation of eigenmodes defines the name of the entire method regardless the approach used for computation of expansion coefficients, we use full abbreviations such as IPEMM, ITEMM, etc. All computational time test calculations in the paper besides the last example have been performed on an Intel Xeon E5-2680 (2.7 GHz) processor using only one core without any parallelism i.e. in sequential mode. As reference in this test we adopt the PSMM method from paper [18] where $N=2M$. This assures adequate accuracy of calculated eigenvalues approximately comparable to that of the other presented methods. As direct eigenvalue solver the LAPACK subroutine ZGEEVX [36] from Intel MKL library is implemented.

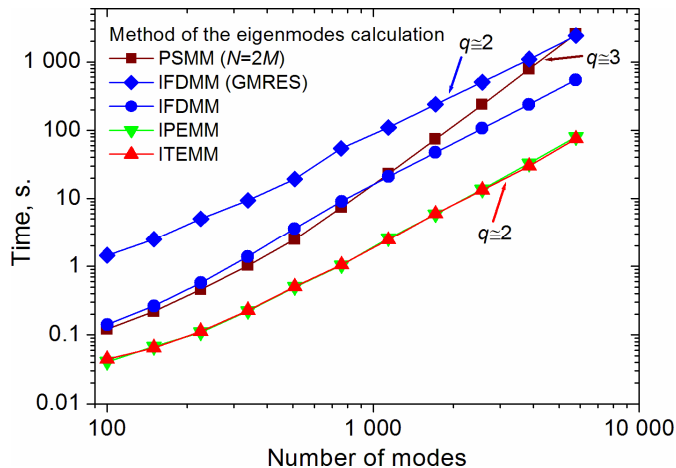

Fig. 2. Computational time of the eigenmodes calculation by various methods for Layer 1 of the metallic grating of Fig. 1a). The quantity $q$ characterizes the slope of the curve.

Fig. 2 shows the relatively higher performance of the iterative method IPEMM compared to its direct PSMM counterpart [18] that exhibits comparable accuracy. When the number of modes is about 6000 the calculation time $t$ differs by a factor higher than 32. Similarly it can be observed for the other methods under consideration. The direct version of ITEMM featuring the matrix size $2M\times2M$ would require the same computational time of PSMM. The matrix size of direct counterpart of IFDMM would be $16M \times 16M$ and the difference in computation time is expected to be greater.

Moreover, a further performance enhancement of IPEMM and ITEMM methods by a factor up to 2 is possible if each part of eigenvalues is calculated by using the minimum possible number of mesh points. Nevertheless, for simplicity of programming we do not implement this technique and calculate the spectrum with a fixed mesh adopting as number of points $N=2M$.

The symbol $q$ in Fig. 2 characterizes the slope of the curve marking the order of computational complexity: $q \sim \ln t / \ln M$. For a small number of modes the number $q$ is slightly smaller than 3 and 2 in case of direct (PSMM) and iterative (IPEMM, ITEMM) methods, respectively. Probably this can be ascribed to the contribution of the time required to move data in computer memory which is proportional to the second power of the number of modes for direct and to the first power for iterative algorithms. However, the relative weight of this contribution becomes negligible with increasing number of modes.

The IFDMM [10] adopting the mesh with number of points $N=16M$, solves Eq. (11) by direct and by iterative GMRES solver. In addition it finds eigenmodes part by part and exhibits order of complexity $q=2$ dominating over the direct PSMM solver when large amount of modes are calculated.

Fig. 3 shows the computational time of the expansion coefficients calculation, the stage following the eigenmodes calculation, versus the number of modes for the considered metallic grating. The eigenmodes for all illustrated approaches are calculated within IPEMM method.

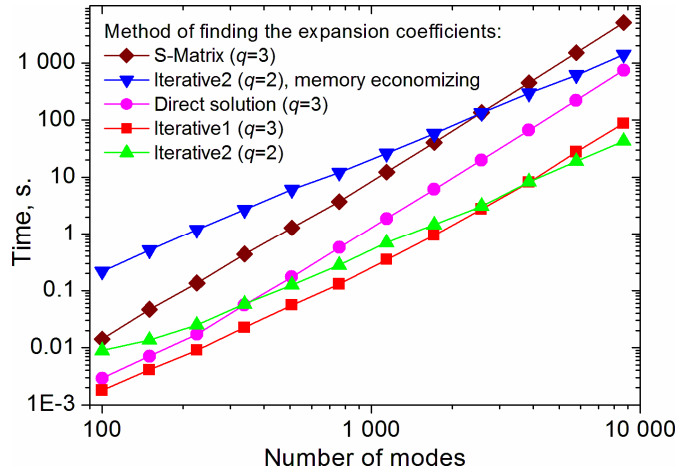

Fig. 3. Computational time of the expansion coefficients calculation versus number of modes for the metallic grating of Fig. 1a). The quantity $q$ characterizes the slope of the curve.

The S-matrix, the direct solver of the Eq. (22) and the two discussed iterative approaches are compared; in particular "Iterative1" and "Iterative2" algorithms use the preconditioners defined by Eq. (26) and by Eq. (29), respectively. The "Iterative2" approach is represented also by its memory economizing modification that does not store the appropriate matrix in the memory but permanently calculates its elements through the iteration process. The precision of finding the coefficients by GMRES subprogram is $10^{-12}$.

It can be seen that all iterative methods outperform S-matrix algorithm, including the memory economizing "Iterative2" method that exhibits lower computational time than S-matrix when the number of modes become greater than 2500. The "Iterative1" algorithm consumes only $2/3 M^3$ flops per layer and is the fastest one for number of modes below four thousands, above the algorithm "Iterative2" exhibiting computational complexity $q=2$ outperforms "Iterative1". The average number of GMRES iteration in this test is equal to 7 for the "Iterative1" algorithm and 67 for the "Iterative2".

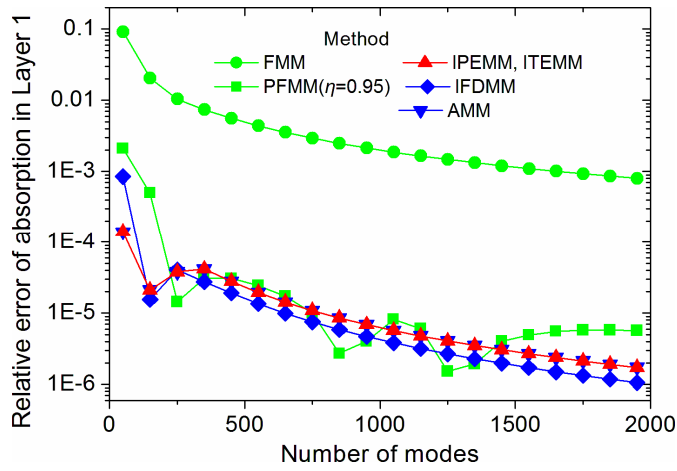

Fig. 4. Relative error of the absorption in Layer 1 of the metallic grating of Fig. 1a).

Fig. 4 shows the relative accuracy of the absorption in Layer 1 (the protrusion) of the discussed grating versus number of the expansion modes obtained by means of various methods in the TM case. As the reference value we exploit 0.03810639822 calculated by the technique discussed in the last example of this section. In case of iterative methods, the dependence of the relative accuracy versus number of modes is not smooth for metallic grating and to plot Fig. 4 we use median of values calculated over the quasiperiod of oscillation of four neighboring points (eigenmodes). It is worth noting that the calculation of the average also produces an accurate result. We use this procedure here for all methods excluding the FMM and PFMM. Such procedure is not particularly time consuming because for the methods based on exact eigenmodes or their approximations we do not need to determine all already calculated eigenmodes when $M$ changes, differently to methods such as FMM for which the complete set of eigenmodes of the corresponding matrix has to be recalculated. After finding the $M$ eigenmodes we calculate absorption by using $M,...,M$-3 of them, by computing for each the corresponding expansion coefficients.

Due to Gibbs phenomenon in grating materials with large permittivity contrast, in this case FMM is outperformed by iterative methods based on spectral elements as well as by IFDMM. Adaptive coordinate transformation in PFMM strongly reduces such effect.

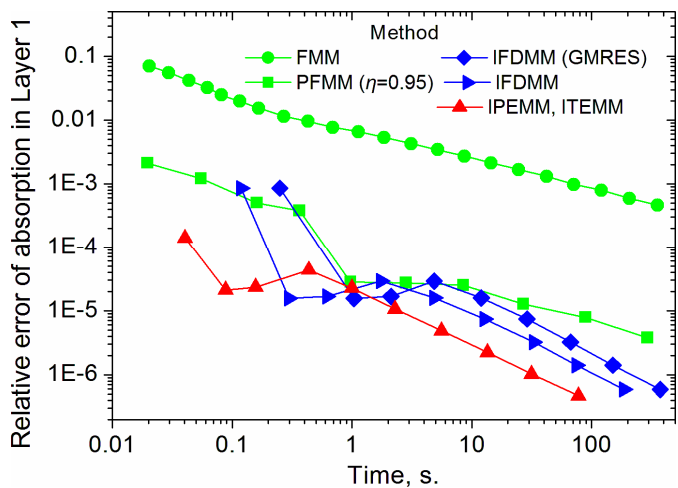

Fig. 5. Relative error of the absorption in Layer 1 of the metallic grating of Fig. 1a) as function of calculation time.

Fig. 5 illustrates relative accuracy of the absorption in Layer 1 of the discussed grating as function of the average computational time. In all time tests presented in the paper we do not reuse any data from previously calculated points besides calculating medians or averages as pointed out above. In addition, the calculation time tests have been repeated up to 100 times making the standard deviation negligible. In this example the "Iterative1" method is used in combination with all methods with the exception of FMM and PFMM with which the S-matrix approach is implemented. It can be seen from Fig. 5 that all presented iterative methods outperform FMM and PSMM in this test.

The next example is the investigation of the relative accuracy of the transmittance of the dielectric grating sketched in Fig. 1a) in which $\varepsilon_1=(2.3)^2$, $\varepsilon_2=(1.5)^2$, $h=1$ $\mu$m, $p=2$ $\mu$m, $d=0.234p$ under the TM polarized light with $\lambda=1$ $\mu$m and $\theta=30^0$. This grating has been previously examined in [10,13,17]. As reference value we use transmitted first order $T_1=0.510592363200$.

From Fig. 6 we observe that the curves of accuracy versus number of modes of all considered methods besides PFMM are almost the same in this case.

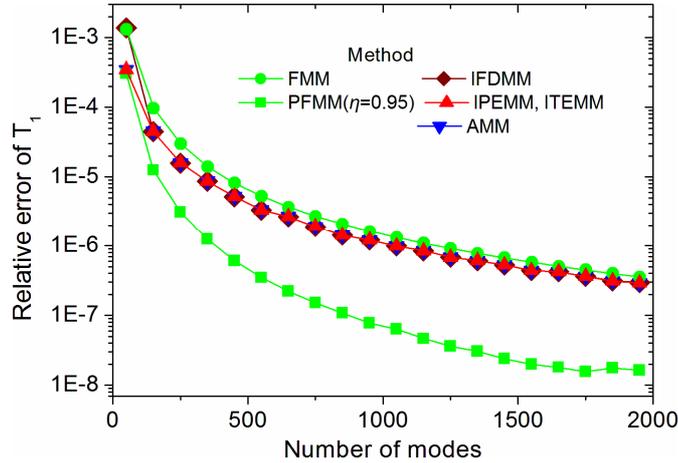

Fig. 6. Relative error of the transmitted first order $T_1$ for the dielectric grating of Fig. 1a).

However, the picture dramatically changes when the methods are examined in time-accuracy coordinates (Fig. 7).

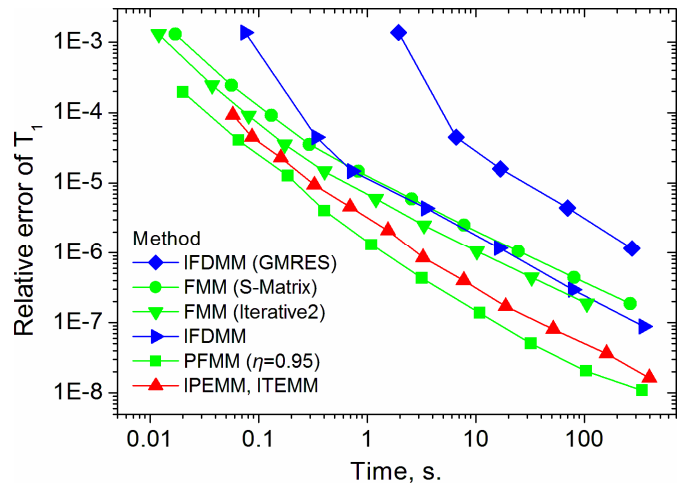

Fig. 7. Relative error of the transmitted first order $T_1$ for the dielectric grating of Fig. 1a) as function of calculation time.

As in the previous test, also in this case the "Iterative1" ($q=3$) method is used within all methods besides PFMM and FMM, where the S-matrix as well as "Iterative2" approaches are implemented. The IPEMM and ITEMM methods outperform FMM with S-matrix technique by a factor higher than 10 when small relative error is required. The IFDMM method with iterative implementation of shift-and-invert technique exhibits higher relative error (Fig. 7) for a given computation time with respect to FMM in this case, but due to steeper slope of the curve is expected to exhibit relatively better performance in case of larger number of modes used. We observe, that application of the "Iterative2" approach to FMM allows a speed up by a factor higher than 2. The PFMM is more exact in this test. Nevertheless, it is very sensitive to precision of eigenmodes calculation. That often leads to dramatic loss of accuracy when a large number of modes is used.

As next test example we exploit a complex periodic symmetric structure Fig. 1b) similar to that discussed in [37]. The incident TM polarized plane wave has wavelength $\lambda=0.75$ $\mu$m with the angle of incidence $\theta=30^0$. The permittivities of materials are $\varepsilon_1=(1.61+0.015i)^2$, $\varepsilon_2=(4.07+0.23i)^2$ and $\varepsilon_3=(0.15+4.9i)^2$ (correspond to Indium Tin Oxide (ITO), amorphous silicon and silver, respectively). The geometrical features are: $h_l=0.1\mu m$ ($l=1,2,4,5$), $h_3=0.3\mu m$, $d_1=0.6\mu m$, $d_2=0.4\mu m$, $d_3=0.2\mu m$ and period $p=1\mu m$.

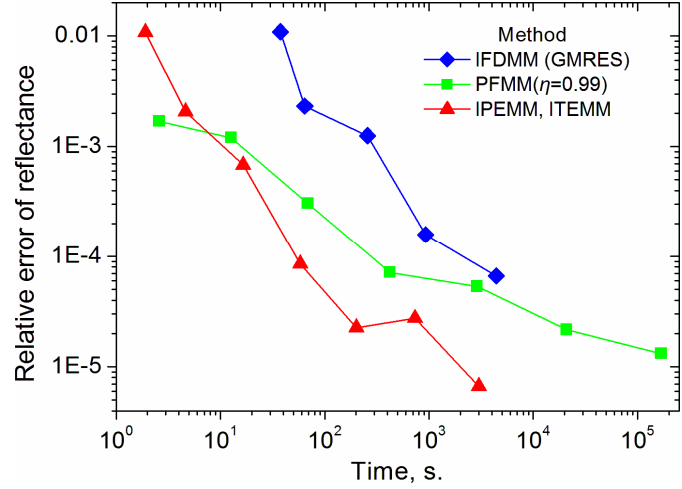

Fig. 8. Relative error of the reflectance of the grating of Fig. 1b) as function of calculation time.

Fig. 8 shows the relative accuracy of the reflectance versus calculation time. The points on the graphs correspond to number of the expansion modes in accordance with relation: $100 \times 2^n + 1$, $n=0,1,2\ldots$ For all methods we calculate the mean of values of reflectance at 10 neighboring eigenmodes similarly to the first example. The "Iterative 2" ($q=2$) algorithm is used with all methods besides PFMM. As reference we consider the value of reflectance 0.0503309. It can be seen from Fig. 8 that the reflectance in case of PFMM exhibits a slower convergence than that observed in case of iterative methods.

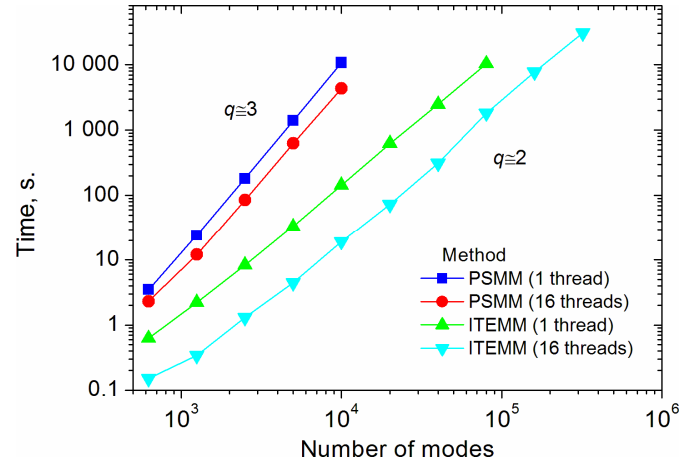

Fig. 9. Computational time of the eigenmodes calculation for Layer 1 of metallic grating of Fig. 1a). The number of threads of parallel execution is writen in parenthesis. The quantity $q$ characterizes the slope of the curve.

An additional useful feature of the presented iterative approach is the possibility to perform calculations with a significantly large number of modes. This allows to use these methods as benchmark. As an example the calculation of zero reflective order of the same metallic grating as in the first test is considered. We performed the calculations on dual-processor system with two Intel Xeon E5-2690 (2.9 GHz) processors in parallel computation, Fig. 9. For comparison data related to sequential processing are also presented. The number of mesh nodes $N$ is equal to $2M$. The ratio of calculation times in case of PSMM with $2M \times 2M$ matrix to that required by ITEMM at $M=10^4$ is approximately 226. Furthermore, it can be seen that the calculation of the eigenvalues by parts is much better parallelizable.

In Fig. 10 the relative accuracy of the zero reflected order $R_0$ for TM case is illustrated. As reference value we use $R_0$=0.848481678905. Calculations were carried out with up to $M_{max}$=3.2×10$^5$ eigenmodes. In this case the general S-matrix algorithm would require about 6 Tb of computer memory only to keep one S-matrix in complex double precision format. The memory economizing "Iterative2" algorithm uses only few gigabytes of memory to store necessary data and can be executed on conventional PC.

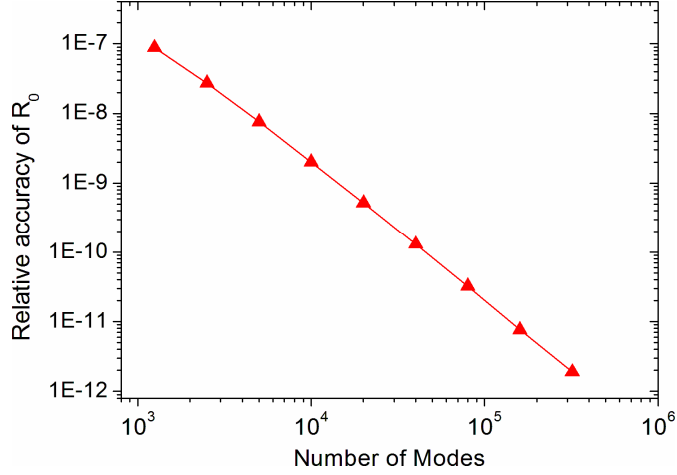

Fig. 10. Relative error of the zero-order reflected efficiency $R_0$ for the metallic grating of Fig. 1a) with $\varepsilon_1=\varepsilon_2=(0.22+6.71i)^2$, $\lambda$=1 μm and $\theta$=30$^0$ calculated for the TM case.

Within this test the eigenvalues obtained by ITEMM solver were additionally purified by Newton's method to relative accuracy better than 10$^{-16}$ by using quad-precision arithmetic, as discussed in [25]. Such purification takes a fraction of a percent of the total computation time considerably raising reliability of the results when $M$ is of order of hundreds thousands. The reference value is found by extrapolating the results calculated by using few points $M$ to $M=\infty$.

## 4. Conclusions

In this paper the possibility to decrease the computation complexity of modal methods, i.e. methods based on eigenmodes expansion, from the third power to the second power of the number of modes is discussed. The main novel parts behind the proposed approach are: calculation of eigenmodes by parts by implementing shift-and-invert iterative technique and adoption of iterative methods to solve the linear equations for the computation of the expansion coefficients. Different implementations of modal methods using expansions by polynomials, trigonometric functions and finite-difference approach that practically exploit these ideas have been demonstrated. We show that the implementation of iterative approach to calculate eigenmodes and the modal expansion coefficients can significantly increase the speed of the method.

In our opinion, the iterative approach is potentially more complicated to program compared with the direct one. Nevertheless, it is very powerful when calculation with large number of modes is required. Moreover, the discussed iterative techniques of calculating expansion coefficients combined with any conventional direct method of eigenmodes computation such as FMM allow to markedly reduce the calculation time.

## Appendix A

The solution of Eq. (11) with zero boundary conditions can be found in form of Chebyshev polynomials expansion:

$$\psi_j^0(x) \equiv \varphi(\xi) = \sum_{n=2}^{N^j} d_n \left[T_n(\xi) - T_{n-2}(\xi)\right], \tag{32}$$

where by means of the following change of variable:

$$\xi = -1 + 2\left(x - x_0^j\right)/L^j = -1 + \gamma^j\left(x - x_0^j\right) \tag{33}$$

the *j*-th element is mapped within the interval (-1,1). Let us remind that $(N^j - 1)$ equals to the number of inner mesh points within *j*-th element. In the following, we omit the index *j* for conciseness. By reordering the sum in Eq. (32):

$$\varphi = \sum_{n=0}^{N}(d_n - d_{n+2})T_n(\xi), \quad d_0 = d_1 = d_{N+1} = d_{N+2} = 0, \tag{34}$$

and by substituting it to Eq. (11), the equation to solve can be rearranged as:

$$\sum_{n=0}^{N-2}\left[\gamma^2 d_n^{(2)}/k_0^2 + (\varepsilon - \kappa)(d_n - d_{n+2})\right]T_n(\xi) = \sum_{n=0}^{N-2} f_n T_n(\xi), \tag{35}$$

where expansion coefficients $d_n^{(2)}$ of the second derivative of $\varphi$ will be found at a later step and expansion coefficients $f_n$ of function in the right side of (11) can be calculated by using Fast Fourier Transform [32]. Eq. (35) implies:

$$\gamma^2 d_n^{(2)}/k_0^2 + (\varepsilon - \kappa)(d_n - d_{n+2}) = f_n, \quad n = 0..N-2 \tag{36}$$

The expansion coefficients $d_n^{(1)}$, $d_n^{(2)}$ of the first and second derivative of $\varphi$ can be determined through relations [32]:

$$2T_n(\xi) = a_n \frac{T'_{n+1}(\xi)}{n+1} - \frac{T'_{n-1}(\xi)}{n-1}, \quad a_n = \begin{cases} 2, n = 0 \\ 1, n \geq 1 \end{cases}, \tag{37}$$

where we pose that $T_{-1} \equiv 0$, $T'_0 = 0$. Since the derivative of $\varphi$ is:

$$\varphi'(\xi) = \sum_{n=1}^{N}(d_n - d_{n+2})T'_n(\xi), \tag{38}$$

and it is possible to write:

$$\varphi'(\xi) = \sum_{n=0}^{N-1} d_n^{(1)} T_n(\xi) = \sum_{n=0}^{N-1} \frac{d_n^{(1)}}{2}\left[a_n \frac{T'_{n+1}(\xi)}{n+1} - \frac{T'_{n-1}(\xi)}{n-1}\right],$$
$$= \sum_{n=1}^{N}\left(a_{n-1}d_{n-1}^{(1)} - d_{n+1}^{(1)}\right)T'_n(\xi)/(2n), \quad d_N^{(1)} = d_{N+1}^{(1)} = 0 \tag{39}$$

consequently:

$$a_n d_n^{(1)} = d_{n+2}^{(1)} + 2(n+1)(d_{n+1} - d_{n+3}), \quad n = 0..N-1$$
$$d_n^{(1)} = 0, n \geq N \tag{40}$$

If $d_n$ are known then all $d_n^{(1)}$ can be found from (40) in descend order. The derivatives at the end points of element in Eq. (14) can be obtained as:

$$\varphi'(-1) = \sum_{n=0}^{N-1} d_n^{(1)} T_n(-1) = \sum_{n=0}^{N-1}(-1)^n d_n^{(1)},$$
$$\varphi'(1) = \sum_{n=0}^{N-1} d_n^{(1)} T_n(1) = \sum_{n=0}^{N-1} d_n^{(1)} \tag{41}$$

and $(\psi_j^0)' = \gamma\varphi'$. The equations for second derivative are calculated similarly:

$$\varphi''(\xi) = \sum_{n=2}^{N}(d_n - d_{n+2})T''_n(\xi), \tag{42}$$

and

$$\varphi'' = \sum_{n=0}^{N-2} d_n^{(2)} T_n(\xi) = \sum_{n=0}^{N-1} \frac{d_n^{(2)}}{2}\left[\frac{a_n T'_{n+1}(\xi)}{n+1} - \frac{T'_{n-1}(\xi)}{n-1}\right]$$
$$= \frac{1}{4}\sum_{n=2}^{N}\left[\frac{a_{n-2}d_{n-2}^{(2)}}{n(n-1)} - \frac{2d_n^{(2)}}{(n+1)(n-1)} + \frac{d_{n+2}^{(2)}}{n(n+1)}\right]T''_n(\xi) \tag{43}$$

Therefore,

$$\frac{a_{n-2}d_{n-2}^{(2)}}{n(n-1)} - \frac{2d_n^{(2)}}{(n+1)(n-1)} + \frac{d_{n+2}^{(2)}}{n(n+1)} = 4(d_n - d_{n+2}). \tag{44}$$

Eq. (36) and (44) represent a system of linear equations with band matrix that can be solved by order of $N$ arithmetic operations. The values of function $\psi_j^0$ in mesh points are founded by using FFT.

**Appendix B**

The solution of Eq. (11) with zero boundary conditions can alternatively be represented by using a sine series within the interval $[0,\pi]$:

$$\begin{aligned}\psi_j^0(x) &\equiv \varphi(\xi) \leftrightarrow \sum_n d_n \sin(n\xi) \\ \xi &= \pi\left(x - x_0^j\right)/L^j = \gamma\left(x - x_0^j\right) \in [0,\pi]\end{aligned}, \tag{45}$$

However, this series alone is not enough to satisfy the right hand side $f$ in (11), which in the most general case has nonzero values at the edges of $j$-th element. In addition, all even derivatives of result of $(\Lambda - \kappa)$ operator applied to (45) also assume values equal to zero on boundaries. If $f$ behaves in another way, then the series will converge slowly. To satisfy these requirements an auxiliary function $\vartheta$ must be added to the series (45):

$$\varphi(\xi) = \vartheta(\xi) + \sum_{n=1}^{N-1} d_n \sin(n\xi). \tag{46}$$

The function $\vartheta$ can be chosen, for instance, as a short series:

$$\vartheta(\xi) = \sum_{n=1}^{K} a_n \left[\cos(n+1)\xi - \cos(n-1)\xi\right], \tag{47}$$

with a number of terms $K$ much smaller than $N$. The number $K$ is chosen depending upon the required speed of convergence (46) to the exact result. It can be shown that the algebraic index of convergence of the series (46) will be $(2K+2)$ if coefficients $a_n$ are chosen so that results of operator $(\Lambda - \kappa)$ applied to (47) and their even derivatives manifest the same behavior as the right hand side $f$ at the edges of element:

$$\begin{aligned}(\Lambda - \kappa)\vartheta\big|_{\xi=0} &= f(0), \quad (\Lambda - \kappa)\vartheta\big|_{\xi=\pi} = f(\pi) \\ \left[(\Lambda - \kappa)\vartheta\right]^{(2)}\big|_{\xi=0} &= f^{(2)}(0), \quad \left[(\Lambda - \kappa)\vartheta\right]^{(2)}\big|_{\xi=\pi} = f^{(2)}(\pi).\end{aligned} \tag{48}$$

…

The left sides of equations (48) can be expressed analytically and the derivatives of $f$ at boundaries can be found for example by finite differences. To achieve the sixth order of convergence in (46) only two terms in (47) are enough and a system of four linear equations must be solved. In this work we mainly employ fourth order of convergence.

By subtracting the result of operator $(\Lambda - \kappa)$ applied to (47) from right hand side $f$, we obtain: $g(\xi) = f(\xi) - (\Lambda - \kappa)\vartheta(\xi)$. By applying the operator $(\Lambda - \kappa)$ to remaining sine series (46) and equated to $g(\xi)$ the coefficients $d_n$ can be found:

$$\sum_{n=1}^{N-1} \left(-\gamma^2 n^2 / k_0^2 + \varepsilon - \kappa\right) d_n \sin(n\xi) = \sum_{n=1}^{N-1} g_n \sin(n\xi). \tag{49}$$

Thus, $d_n = g_n / \left(-\gamma^2 n^2 / k_0^2 + \varepsilon_j - \kappa\right)$. To decompose $g(\xi)$ in sine series and obtain function values on mesh points from coefficients $d_n$, the FFT can be implemented.

**Acknowledgments**
I.S. is grateful to Taiichi Otsuji, Victor Ryzhii and Akira Satou for valuable discussions.